\documentclass[12pt]{article}
\usepackage{amsmath,amssymb}
\newcommand{\be}{\begin{equation}}
\newcommand{\ee}{\end{equation}}
\newcommand{\bea}{\begin{eqnarray}}
\newcommand{\eea}{\end{eqnarray}}
\newcommand{\ba}{\begin{array}}
\newcommand{\ea}{\end{array}}
\newcommand{\nn}{\nonumber}
\newcommand{\p}[1]{(\ref{#1})}

\renewcommand{\a}{\alpha}  \renewcommand{\b}{\beta}
\renewcommand{\th}{\theta} \newcommand{\bth}{\bar\th}
\newcommand{\ve}{\varepsilon} \newcommand{\bve}{\bar\ve}
\def\bpsi{\bar\psi} \def\bxi{\bar\xi}

\newcommand{\bD}{\bar D}
\newcommand{\Z}{Z} \newcommand{\bZ}{\bar{\Z}}

\newcommand{\bz}{\bar z}
\def\F{{\tt{F}}}
\def\bF{\bar{\F}}
\renewcommand{\H}{{\tt H}}
\newcommand{\bH}{\bar \H}
\newcommand{\Hz}{\H,_z}
\newcommand{\bHz}{\bH,_{\bz}}
\newcommand{\M}{\,{\tt M}\,}
\newcommand{\N}{\,{\tt N}\,}
\newcommand{\tPhi}{\tilde\Phi}

\def\ReA{{\tt Re\,}A}
\def\ImA{{\tt Im\,}A}
\newcommand{\diff}{\partial}

\begin{document}
\thispagestyle{empty}
\vspace{3cm}
\begin{flushright}
\end{flushright}
\vspace{3cm}
\begin{center}
{\Large \bf N=4, d=3 nonlinear electrodynamics}
\end{center}

\vspace{1cm}

\begin{center}
{\large \bf S.~Bellucci${}^{a}$, S.~Krivonos${}^{b}$, A.~Shcherbakov${}^{b}$ }
\vspace{0.5cm}

${}^a$ \textit{INFN-Laboratori Nazionali di Frascati, Via Enrico Fermi 40,
00044 Frascati, Italy}\\
\vspace{0.2cm}

\texttt{bellucci@lnf.infn.it} \vspace{0.4cm}

${}^b$ \textit{Bogoliubov Laboratory of Theoretical Physics, JINR, 141980
Dubna, Russia}\\
\vspace{0.2cm}

\texttt{krivonos, shcherb@theor.jinr.ru}
\end{center}

\vspace{2cm}
\begin{abstract}
We construct a new off-shell $\mathcal{N}{=}4$, $d{=}3$ nonlinear
vector supermultiplet. The irreducibility constraints for the superfields
leave in this supermultiplet the same component content as in the ordinary
linear vector supermultiplet. We present the most
general sigma-model type action for the  $\mathcal{N}{=}4$, $d{=}3$ electrodynamics
with the nonlinear vector supermultiplet, which
despite the nonlinearity of the supermultiplet may be written as an integral over a chiral
superspace. This action share the most important properties with its linear counterpart.
We also perform the dualization of the vector component into a scalar one and find
the corresponding $\mathcal{N}{=}4$, $d{=}3$ supersymmetric action which describes new hyper-K\"ahler
sigma-model in the bosonic sector.
\end{abstract}

\newpage \setcounter{page}{1}

\section{Introduction}
During the last few years it has been clarified that the nonlinear supermultiplets play
an exceptional role in supersymmetric mechanics with extended $N=4,8$ supersymmetries.
In contrast to higher dimensional theories, where extended supersymmetries imply K\"ahler
or hyper-K\"ahler geometries in the bosonic sigma-model parts, the constraints on the geometry
in one-dimensional cases admit another type of geometries - hyper-K\"ahler or octonionic K\"ahler
geometries with torsion  \cite{GPS} - \cite{N48a}. Moreover, only such sigma-models appear when
considering linear supermultiplets. Thus, in order to describe another geometries
in extended supersymmetric mechanics we are obliged to involve the nonlinear supermultiplets.

Despite the fact that many new off-shell nonlinear one-dimensional supermultiplets have been
explicitly constructed in superfield and component approaches in \cite{bbks}-\cite{nl2}, it is still
unclear whether there are any higher dimensional analogs of such supermultiplets. In the present paper
we are going to demonstrate that at least one of the $N=8,d=1$ nonlinear supermultiplets has its $N=4, d=3$
counterpart. In  Section 2 we present the corresponding superfield constraints defining a new
nonlinear vector multiplet and demonstrate that it possesses the same component structure as its
linear version. When dealing with the nonlinear supermultiplets the most serious problem is to find
a proper invariant action for them. Fortunately, for the new nonlinear vector supermultiplet the
action has the same form as for the linear one. So, the role of nonlinearities is to deform the geometry
of the bosonic part and the interaction terms. In Section 3 we analyse the component structure of the
action for the nonlinear supermultiplet. In full analogy with the linear supermultiplet
there is a gauge field among the components
of nonlinear supermultiplets,  which appears  through its field-strength only.
Yet, in three dimensions the gauge field is dual to a scalar. This suggests the existence of another $N=4,d=3$
supermultiplet which contains only four scalars in its bosonic sector. We construct
such a multiplet and explicitly demonstrate that the bosonic geometry is of the hyper-K\"ahler type, as it should be.
The observed similarity of the proposed nonlinear vector supermultiplet  to the
standard linear case raised the question which supermultiplet is preferable. We shortly discuss this situation in the
Conclusion.

\section{N=4, d=3 nonlinear vector supermultiplet}
In this Section we perform the direct construction of the nonlinear vector supermultiplet and analyse
its component structure.

The key idea of our construction comes from the superfield description of the $N=8,d=1$ supermultiplet $(2,8,6)$ \cite{bbks}.
Indeed, the constraints defining this supermultiplet may be easily lifted up to three dimensions. So, let us define
our nonlinear supermultiplet by the modified nonlinear chirality constraints
\be\label{nlchir}
D^i_\a \bZ = - \lambda \bZ \bar D^i_\a \bZ, \qquad  \bar D_{i\,\a} \Z = \lambda \Z D_{i\,\a} \Z, \qquad \lambda=const,
\ee
together with the linear reality conditions
\be\label{irred}
D^\a_i D_{j\a} \Z = \bar D_{i\a} \bar D^\a_j \bZ.
\ee
Here, $\lambda$ is a constant parameter, the scalar complex superfield $\Z$ lives in a three-dimensional space
$\{ x^{\a\b}\} \in \mathbf{R}^3$
augmented by eight Grassmann variables $\th^i_\a$, $\bth_{i\a}$ and the covariant spinor derivatives satisfy the
following relations:
\be\label{salg}
\{ D^i_\a\,, \bar D_{j\b} \} = i \delta^i_j \, \partial_{\a\b},
\ee
with $\a,\b=1,2$ being the $d=3$ $sl(2,R)$ spinor indices, while  $i,j=1,2$ refer to the doublet indices of the $SU(2)$
automorphism group.

Before analyzing the component structure of our supermultiplet, let us note that with a natural dimension of the
superfield $\left[ \Z \right]=cm^{-1/2}$ the parameter $\lambda$ has also a nontrivial dimension $\left[ \lambda \right]=cm^{1/2}$.
Nevertheless, it is preferable to consider a dimensionless superfield $\Z$ and, therefore, a dimensionless parameter $\lambda$.
Another point concerns the value of the parameter $\lambda$. If it does not vanish, it is always
possible to pass to $\lambda=1$ by a redefinition of the superfields $\Z, \bZ$.
So, we are left with only two essential values $\lambda=1$ and $\lambda=0$. In what follows we
will put $\lambda=1$.
The proper, natural dimensions of the superfields and the $\lambda$-dependence may  be easily restored, if needed.

With $\lambda=0$ the constraints \p{nlchir},\p{irred} describe the linear $N=4, d=3$ vector supermultiplet.
The component structure of the $N=4$ superfield $\Z$, implied by \p{nlchir}, \p{irred}, is a bit involved in comparison
with the $\lambda=0$ case. In order to define it, let us firstly consider the constraints \p{nlchir}. It
immediately follows from \p{nlchir} that the derivatives ${\bar D}_{i\,\a}$  of the superfield $\Z$ (or $D^i_\a$
of $\bZ$) can be expressed as $D^i_\a$ (or ${\bar D}_{i\,\a}$) derivatives of the same
superfield. Therefore, as in the case of linear chirality  constraints, only the components appearing in
$\theta^{i\,\a}$-expansion of $\Z$ and  $\bar\theta_{i\a}$-expansion of the  superfield $\bZ$ are independent.
Let us define these components as follows:
\be\label{compdef}
\ba{l}
z=\Z,\\
\psi^i_\a = D^i_\a \Z, \\
A_{\a\b} = D^i_\a D_{i\b}\Z,\\
B^{ij} = D^{i\a} D^j_\a\Z, \\
\xi^i_\a = D^{i\b} D^j_\b D_{j\a}\Z, \\
X = D^{i\a} D^j_\a D^\b_i D_{j\b} \Z,
\ea \qquad
\ba{l}
\bar z = \bZ,\\
 \bpsi_{i\a} = \bar D_{i\a} \bZ,\\
\bar A_{\a\b} = \bar D_{i\a} \bar D^i_\b \bZ, \\
\bar B_{ij} = \bD_{i\a} \bD_j^\a \bZ, \\
\bxi_{i\a} = \bar D_i^\b \bar D_{j\b} \bar D_\a^j \bZ,\\
\bar X = \bar D_{i\a} \bar D_j^\a \bar D_\b^i \bar D^{j\b} \bZ.
\ea
\ee
The right hand side of each equation is supposed to be taken upon $\theta^{i\,\a}=\bar\theta_{i}^{\a}=0$.

Thus, the first part of
our constraints \p{nlchir} leaves in the $N=4, d=3$ superfields $\Z, \bZ$ sixteen bosonic and sixteen fermionic components, like
in the case of the liner supermultiplet.

Now it is time to consider the constraints \p{irred}. In terms of the components \p{compdef} the constraints (\ref{irred}) acquire the following form:
\bea\label{comp2}
B_{ij} &=& \bar B_{ij},\nn\\
\xi^i_\a &=& -3i\partial_{\a\b}\bpsi^{i\b} - \frac 32 \bar A_{\a\b}\bpsi^{i\b} - \frac 32 \bar B^{ij}\bpsi_{j\a} - \bar z \bxi^i_\a,\nn\\
\bxi_{i\a} &=& -3i\partial_{\a\b}\psi^\b_i + \frac 32 A_{\a\b}\psi^\b_i - \frac 32 B_{ij}\psi^j_\a + z \xi_{i\a},\nn\\
X &=& -6 \partial^{\a\b}\partial_{\a\b} \bar z + 6i\partial_{\a\b} \left( \bpsi^\a_i \bpsi^{i\b} + \bar z \bar A^{\a\b}\right)
    - 6 \bar B_{ij} \bpsi^i_\a\bpsi^{j\a} + 6 \bar A^{\a\b} \bpsi_{i\a}\bpsi^i_\b +\nn\\
    &+& 3 \bar z \bar A_{\a\b} \bar A^{\a\b} - 3 \bar z \bar B_{ij} \bar B^{ij} - 8\bar z \bpsi_{i\a} \bxi^{i\a} - \bar z^2 \bar X,\\
\bar X &=& -6\partial^{\a\b}\partial_{\a\b} z - 6i\partial_{\a\b} \left( \psi^{i\a} \psi^\b_i + z A^{\a\b}\right)
    - 6 B_{ij} \psi^{i\a}\psi^j_\a + 6 A^{\a\b} \psi^i_\a \psi_{i\b}+ \nn\\
    &+&3 z A^{\a\b} A_{\a\b} - 3 z B^{ij} B_{ij} + 8 z \psi^{i\a} \xi_{i\a} - z^2 X,\nn
\eea
and
\bea\label{A2}
&& \diff_{\a\b} \left[ (1 + z \bz) (A^{\a\b} + \bar A^{\a\b} ) + 2i (z\diff^{\a\b} \bz - \bz \diff^{\a\b} z)
    + 4 \psi^{i(\a} \bpsi^{\b)}_i \right]+\nn\\
&& \phantom{\diff_{\a\b}x}\frac i2 (A_{\a\b} - \bar A_{\a\b} ) \left[ (1 + z \bz) (A^{\a\b} + \bar A^{\a\b} )
    + 2i (z\diff^{\a\b} \bz - \bz \diff^{\a\b} z) + 4 \psi^{i(\a} \bpsi^{\b)}_i \right] = 0,\label{A}\\
&& \diff_{\a(\b}(A_{\mu)}{}^\a - \bar A_{\mu)}{}^\a) = 0.\nn
\eea
The equations  \p{comp2} impose the reality condition on $B_{ij}$ and express the higher bosonic components $X, \bar X$ and fermionic
 $\xi^i_\alpha, \bar\xi_{i\alpha}$ ones in terms of physical bosons and fermions and
auxiliary bosons $B_{ij}, A^{\a\b}$. For the auxiliary bosons $A^{\a\b}$ we have the differential equations \p{A2}.
In the $\lambda=0$ case, which corresponds to discarding all nonlinear terms, these equations read
\be\label{lin}
\diff_{\a\b} (A^{\a\b} + \bar A^{\a\b} )=0, \quad \diff_{\a(\b}(A_{\mu)}{}^\a - \bar A_{\mu)}{}^\a) = 0.
\ee
The first of these equations defines $\ReA^{\a\b}$ to be a field-strength of the gauge field, while the second one
is the Bianchi identity which allows us to express $\ImA^{\a\b}$ through a scalar field as follows:
\be\label{ImAduallin}
\ImA_{\a\b} = - \diff_{\a\b} \Phi.
\ee

In the $\lambda \neq 0$ case the second of the equations \p{A2} is the same as in the linear case.
Moreover, despite the nonlinearity of the supersymmetry transformations of the components $A_{\a\b}$
\bea\label{trA}
&&\delta A_{\a\b} = -\frac23 \ve^i_{(\a} \xi_{i\b)} - 2i \bve^{i\mu} \diff_{\mu(\a} \psi_{i\b)}
    - \bve^i_{(\a} \psi^j_{\b)} B_{ij} + \bve^i_{(\a} A_{\b)\mu} \psi^\mu_i
    + \frac23 \bve^i_{(\a} \xi_{i\b)} z, \\
&&\delta \bar A_{\a\b} = \frac23 \bve^i_{(\a} \bxi_{i\b)} + 2i \ve^{i\mu} \diff_{\mu(\a} \bpsi_{i\b)}
    - \ve^i_{(\a} \bpsi^j_{\b)} \bar B_{ij} + \ve^i_{(\a} \bar A_{\b)\mu} \bpsi^\mu_i
    + \frac23 \ve^i_{(\a} \bxi_{i\b)} \bz,
\eea
the imaginary part of $A_{\a\b}$ transforms in a linear manner and, moreover, as a total
derivative
\be\label{trImA}
\delta (A_{\a\b} - \bar A_{\a\b} ) = -2i \, \diff_{\a\b} (\ve_{i\mu} \bpsi^{i\mu} + \bve_{i\mu} \psi^{i\mu}),
\ee
which is quite amazing in three dimensions. This means that in full analogy with the linear case \p{ImAduallin} we can identify
\be\label{ImAdual}
\ImA_{\a\b} = - \diff_{\a\b} \Phi,
\ee
where the new physical bosonic field $\Phi$ transforms with respect to $N=4, d=3$ supersymmetry as
\be\label{trPhi}
\delta \Phi =  \ve_{i\mu} \bpsi^{i\mu} + \bve_{i\mu} \psi^{i\mu}.
\ee
What is much more important is that, with the help of this new bosonic field $\Phi$, the first of the equations in (\ref{A2})
may be rewritten as
\be\label{constr}
\diff_{\a\b} \left[ e^{\Phi}\left( (1 + z \bz) \ReA^{\a\b} + i (z\diff^{\a\b} \bz - \bz \diff^{\a\b} z)
    + 2 \psi^{i(\a} \bpsi^{\b)}_i \right)\right] = 0.
\ee
Clearly, this equation is a non-linear variant of the first of the equations \p{lin}. So, we will
treat this expression as a Bianchi identity  which defines a field strength ${\cal F}{}^{ab}$
\be\label{Bianchi}
\diff_{\a\b} {\cal F}{}^{\a\b} = 0, \qquad {\cal F}{}^{\a\b} = e^{\Phi}\left( (1 + z \bz) \ReA^{\a\b} + i (z\diff^{\a\b} \bz
    - \bz \diff^{\a\b} z) + 2 \psi^{i(\a} \bpsi^{\b)}_i \right).
\ee
Thus, we conclude that the constraints \p{nlchir}, \p{irred} define the $N=4, d=3$ nonlinear vector supermultiplet,
which has the same component structure as the ordinary linear $N=4, d=3$ vector multiplet. The net effect of the
nonlinearity of the basic constraints is a more complicated nonlinear structure of the higher components in the
superfield $\Z$ \p{comp2} and the nonlinearity of the field strength ${\cal F}{}^{\a\b}$ \p{Bianchi}.

\setcounter{equation}0
\section{The Action}
When dealing with nonlinear supermultiplets the most serious problem is the construction of
invariant actions, especially in the case of extended supersymmetries. The source of the problems, clearly,
is a too high dimension of the superspace measure. The way out of this problem is well known -- one should write the
superspace action as an integral over  some invariant subspaces. For the linear $N=4, d=3$ vector supermultiplet such
a subspace is a just the chiral superspace. So, the general sigma-model type action in this case reads
\be\label{act1}
S=\kappa\int d^3x\left[ \int d^4 \theta\;\; \F (\Z) + \int d^4 \bar\theta\;\; \bF (\bZ) \right],
\ee
where  $\F(\Z)$ is an arbitrary holomorphic function. What is rather unexpected is that the action \p{act1} is
still invariant in the case of our nonlinear vector supermultiplet. Usually, if the superfields are not chiral, the integration
over the chiral superspace fails to be invariant under supersymmetry. However, for
the nonlinear vector supermultiplet with the constraints \p{nlchir}, \p{irred}, the action \p{act1} is perfectly invariant
with respect to the full $N=4, d=3$ supersymmetry. Indeed, the supersymmetry transformations of the integrand of, for example,
the first term in \p{act1}, which seems to break supersymmetry, read
\be\label{dok1}
\delta \F(\Z) \sim \bar\epsilon{}^{i\a} \bD_{i\a} \F(\Z) = \bar\epsilon{}^{i\a} \F' \Z D_{i\a} \Z.
\ee
It is evident that the right-hand side can always be represented as a $D$-derivative of a function of $\Z$. Hence,
the variation in \p{dok1} disappears after integration over $d^4 \theta$ and therefore the action \p{act1} is
invariant with respect to the full $N=4$ supersymmetry.

After integrating in \p{act1} over the Grassmann variables  we will get the following component action:
\bea\label{L1}
S &=&\kappa \int d^3 x\left[
     \F^{(4)} \psi_i^\a \psi_{j\a} \psi^\b_i \psi_{j\b}
    + 3\F''' \left(B^{ij}\psi_i^\a\psi_{j\a} - A_{\a\b}\psi^{i\a}\psi^\b_i \right) + \right.\nn\\
    &+& \F'' \left( -\frac 32 A^{\a\b}A_{\a\b} + \frac 32 B^{ij}B_{ij} - 4\psi^{i\a}\xi_{i\a}\right)
    + \F' X +\\
    &+& \bF^{(4)}\bpsi_{i\a} \bpsi_j^\a \bpsi_\b^i \bpsi^{j\b}
    + 3\bF''' \left(\bar B^{ij}\bpsi_{i\a}\bpsi_j^\a - \bar A_{\a\b}\bpsi^\a_i\bpsi^{i\b} \right) + \nn\\
    &+& \left. \bF'' \left( -\frac 32 \bar A_{\a\b} \bar A^{\a\b} + \frac 32 \bar B_{ij} \bar B^{ij} + 4\bpsi_{i\a}\bxi^{i\a}\right)
    + \bF' \bar X\right].\nn
\eea
Clearly, the effect of nonlinearity of the basic constraints \p{nlchir} is hidden inside the definitions of the components
\p{comp2}, \p{Bianchi}, because the component action \p{L1} has the same form as for the linear vector supermultiplet.

For the sake of brevity, in what follows we will be interested in the bosonic part of the action, so that we will drop all terms with fermions
\be\label{action}
S \sim \int d^3 x \left[ \F''\left(-\frac32 A^{\a\b}A_{\a\b} + \frac32 B^{ij} B_{ij} \right)
    + \bF''\left(-\frac32 \bar A^{\a\b}\bar A_{\a\b} + \frac32 \bar B^{ij} \bar B_{ij} \right)
    + \F'X + \bF'\bar X \right].
\ee
In order to reveal the nonlinear structure of the action \p{action}, one should insert the expressions for the higher components
$X, \bar X$ \p{comp2}, use the definition of the field strength \p{Bianchi} and the new bosonic field $\Phi$ \p{ImAdual}, and
exclude the auxiliary fields $B_{ij}$ by their equations of motion. Let us start from auxiliary fields.
On a mass shell the fields $B_{ij}$ are expressed in terms of the fermions only, therefore they disappear in bosonic
limit.
Now, using the explicit expressions for components $X$, $\bar X$ (\ref{comp2}) and rewriting
$A_{\a\b}$ as follows:
\be\label{A3}
A_{\a\b} = \frac{e^{-\Phi} {\cal F}_{\a\b} - i (z \diff_{\a\b} \bz - \bz \diff_{\a\b} z) }{1+z\bz}
    - i \diff_{\a\b}\Phi, \;
\bar A_{\a\b} = \frac{e^{-\Phi} {\cal F}_{\a\b} - i (z \diff_{\a\b} \bz - \bz \diff_{\a\b} z) }{1+z\bz}
    + i \diff_{\a\b}\Phi,
\ee
we finally come to the action
\bea\label{action2}
S & =&  \frac32 \kappa \, \int d^3 x \left\{ g \left[
4 \diff^{\a\b} z \diff_{\a\b} \bz
    + \diff^{\a\b} (z\bz) \diff_{\a\b} (z\bz) - e^{-2\Phi} {\cal F}{}^{\a\b} {\cal F}_{\a\b} +
    (1+z\bz)^2 \diff^{\a\b} \Phi \diff_{\a\b} \Phi \phantom{\frac{A}{B}}\right. \right.\nn\\
&&  \left.  + 2 (1+z\bz) \diff^{\a\b}\Phi \diff_{\a\b} (z\bz)+
4i \frac{e^{-\Phi}{\cal F}{}^{\a\b}}{1-z\bz}\left(z\diff_{\a\b}\bz-\bz\diff_{\a\b} z\right)\right]\nn \\
&& \left. + 2e^{-\Phi} \N {\cal F}{}^{\a\b}\left(  \diff_{\a\b}(z \bz) -
    (1-z\bz) \diff_{\a\b} \Phi\right) \phantom{\frac{A}{B}}\right\},
\eea
with the functions $g(z,\bz)$ and $\N (z,\bz)$ defined as
\be
g(z,\bz)=\frac{1-z\bz}{1+z\bz}\M \equiv \frac{1-z\bz}{1+z\bz}\left( \Hz + \bHz\right) , \qquad \N = -i (\Hz - \bHz).
\ee
Here, in order to make the presentation easier, we introduced the following combinations:
\be\label{H}
\H(z,\bz) = \frac{\F' - z^2 \bF'}{1-z^2\bz^2}, \qquad \bH(z,\bz) = \frac{\bF' - \bz^2 \F'}{1-z^2\bz^2}.
\ee
Thus, we see that the nonlinearity of the basic constraints results in the modification of the bosonic action.
The simplest way to see this effect is to consider the particular choice
\be
\F=\frac{z^2}{2}, \qquad \bF=\frac{\bz^2}{2},
\ee
which corresponds to the free action in case of linear vector supermultiplet. With these $\F, \bF$ the action \p{action2} reads
\bea\label{faction}
S &= &3\kappa \, \int d^3 x  \frac{1-z\bz}{(1+z\bz)^2} \left\{ 4 \diff^{\a\b} z \diff_{\a\b} \bz -
4\frac{( z \diff_{\a\b} \bz -\bz \diff_{\a\b} z)^2}{(1-z\bz)^2} \right. \nn \\
&& \left. +  (1+z\bz)^2 \left[ \diff^{\a\b} Y \diff_{\a\b} Y-\left( e^{-Y} {\cal F}_{\a\b}-
2i \frac{z \diff_{\a\b} \bz -\bz \diff_{\a\b} z}{1-z^2 \bz^2}\right)^2\right]
\right\}
\eea
where
\be
Y \equiv \Phi+ \log(1+z\bz).
\ee
Thus, even the simplest choice of the functions $\F, \bF$  yields a rather complicated action.

Finally, we would like to construct the action for the dual version of our nonlinear supermultiplet. As usual, in order to dualize
the field-strength ${\cal F}_{\a\b}$ into a scalar field, one should insert the constraint \p{Bianchi} into the action \p{action2}
with a Lagrange multiplier $y$. Eliminating then the field-strength ${\cal F}_{\a\b}$, we finally get the action
for the supermultiplet with four scalars
\bea\label{AA}
&&S= \kappa\, \int d^3 x \left\{ \frac{3}{2} \M e^{-2\tPhi} \frac{(1-z\bz)}{1+z\bz}^3\,\left[
4 \diff_{\a\b} \left( \frac{z e^{\tPhi}}{1-z\bz}\right) \,\diff^{\a\b} \left( \frac{\bz e^{\tPhi}}{1-z\bz} \right)
    + e^{2\tPhi} \, \diff_{\a\b} \tPhi \,\diff^{\a\b} \tPhi \right] +\right. \nn \\
&& \left. \frac23 \frac{1+z\bz}{(1-z\bz)^3} \frac{1}{e^{-2\tPhi}\M} \left[
    \diff_{\a\b} y + 3i e^{-\tPhi} \M \frac{1-z\bz}{1+z\bz} \, (\bz \diff_{\a\b} z - z \diff_{\a\b} \bz)
    + \frac32 (1-z\bz)^2 e^{-\tPhi} \N \diff_{\a\b} \tPhi \right]^2 \right\}
\eea
where
\be
\tPhi = \Phi + \log(1-z\bz).
\ee
The explicit form of the action \p{AA} suggests a new set of coordinates $u, \bar u, V, y$
\be\label{ccor}
u = 2z e^{\Phi} \, , \qquad  \bar u = 2 \bz e^{\Phi} \, , \qquad V = e^\Phi (1-z\bz),
\ee
in which it acquires a standard Gibbons-Hawking form \cite{GH}
\be\label{GH}
S = \kappa\, \int d^3 x \left[ g \left( \diff_{\a\b} u \diff^{\a\b}{\bar u} + \diff_{\a\b} V \diff^{\a\b} V\right)
 +  \frac{1}{g} \left( \diff_{\a\b}y +A \diff_{\a\b} u +{\bar A} \diff_{\a\b}{\bar u} + A_V \diff_{\a\b} V\right)^2 \right].
\ee
Here
\be
A=\frac{3}{2}\,i\, \frac{ \M (1-z\bz)^2}{V^2 (1+z\bz)}\bz,\quad {\bar A}=-\frac{3}{2}\,i\, \frac{ \M (1-z\bz)^2}{V^2 (1+z\bz)}z,\quad
A_V=\frac{3}{2}\frac{\N}{V^2} (1-z\bz)^3
\ee
and the metric $g$
\be
g\equiv \frac{3}{2} \frac{\M (1-z\bz)^3}{V^2 (1+z\bz)}
\ee
obeys to
\be
\mbox{ rot } \overrightarrow{A} = -\mbox{ grad  }g, \qquad \overrightarrow{A} =\left( A, \bar A, A_V\right)
\ee
in the coordinates \p{ccor}.

It is evident now that this four dimensional manifold has a vanishing Ricci tensor, and the
action \p{GH} describes a hyper-K\"ahler sigma-model. However, the explicit dependence of the action \p{GH} on the scalars
is quite different from the case of the linear $N=4,d=3$ vector supermultiplet. One of the most interesting features
of the action \p{GH} is the existence only one isometry, whereas, in the case of the linear supermultiplet, there are two
isometries.

\section{Conclusion}
In the present paper we proposed a new nonlinear superfield constraint to describe the $\mathcal{N}{=}4$, $d{=}3$
nonlinear electrodynamics. The new nonlinear vector supermultiplet has the same components structure as its
linear counterpart. Moreover, the most general action for the new supermultiplet may be written as an integral over the
chiral subspace, and thus it depends on an arbitrary holomorphic function, in full analogy with the linear case. The novel
features of the proposed nonlinear electrodynamics show up as the essential nonlinearities, which arise even in the case
of the simplest choice for the action.

One of the most amazing features of the proposed nonlinear vector supermultiplet is its similarity to the
standard linear case. Indeed, the component counting, the appearance of the field strength among the components,
the structure of the action --- all these features are mimicking the corresponding structures for linear electrodynamics.
So, in order to decide which supermultiplet is preferable, one should carry out a more subtle analysis, maybe with some
specific action. In this respect, it seems interesting to perform a more detailed investigation of the
dualized version of the nonlinear $N=4,d=3$ vector supermultiplet. Clearly, after dualization we
deal with a new version of a hypermultiplet. The geometry of the bosonic manifold is of the hyper-K\"ahler type,
but it is different from the one arising after dualization of a linear vector supermultiplet. It is still
unclear whether this new deformed geometry possesses a four-dimensional analog. Another interesting
problem is to construct a new variant of the Chern-Simons action with the nonlinear supermultiplet.

Notice that in the present paper we fixed a unit value
of the parameter $\lambda$. In $\mathcal{N}{=}4$, $d{=}1$ this parameter can be chosen as an arbitrary holomorphic function,
parameterizing a family of supersymmetric extensions of the given system \cite{bn}. The nature of this parameter
in $d=3$ would deserve further investigation.

Finally, one should mention that we have constructed our nonlinear supermultiplet within the framework where
one of the scalars is hidden. It appears only through derivatives inside components. It is known that
there is another formulation of the $N=4,d=3$ vector supermultiplet in the harmonic superspace \cite{bz}, where this scalar
component is manifest. It would be quite worthwhile to accomplish a full analysis of the nonlinear supermultiplet
within this approach.

\end{document}